\documentclass[cits]{PoS}
\usepackage{natbib, graphicx}
\setcitestyle{square}
\setcitestyle{numbers}

\title{Black Hole -- Galaxy Co-evolution}

\ShortTitle{Black Hole -- Galaxy Co-evolution}

\newcommand\aj{{AJ}}
\newcommand\araa{{ARA\&A}}
\newcommand\apj{{ApJ}}
\newcommand\apjl{{ApJ}}

\newcommand\aap{{A\&A}}
\newcommand\nat{{Nature}}
\newcommand\mnras{{MNRAS}}

\newcommand\pasj{{PASJ}}

\def\LOIIIs4{$L[\mbox{O\,{\sc iii}}]$/$\sigma^4$}

\def\Msigma{$M_{\rm BH} - \sigma$}

\def\Msun{$M_{\odot}$}

\def\ergs{$~\rm erg s^{-1}$}

\author{\speaker{Kevin Schawinski}\\
	Einstein Fellow\\
        Department of Physics, Yale University, New Haven, CT 06511, U.S.A.\\
	Yale Center for Astronomy and Astrophysics, Yale University, P.O. Box 208121, New Haven, CT 06520, U.S.A.\\
       E-mail: \email{kevin.schawinski@yale.edu}}

\abstract{The growth of black holes and the formation and evolution of galaxies appear to be linked at such a fundamental level that we think of the two as `co-evolving.' Recent observations show that this co-evolution may be complex and the result of several different pathways.  While it is clear that black hole accretion is linked to specific phases of the evolution of the host galaxy, the impact of the energy liberated by the black hole on the evolutionary trajectory of the host by feedback is less clear. In this contribution, I review the motivations for co-evolution, the current state of the observational picture, and some challenges by black hole feedback. }

\FullConference{Frank N. Bash Symposium New Horizons In Astronomy,\\
		October 9-11, 2011\\
		Austin Texas}

\begin{document}

\section{Why Co-evolution?}

The impact that accreting black holes can have on their surrounding galaxies is profound yet still poorly understood. A full description of why, how, and when black holes alter the evolutionary pathways of their host galaxies remains one of the major outstanding questions in astrophysics.

Semi-analytic models of the formation and evolution of galaxies from cosmological initial conditions cannot produce observed galaxy population properties without the additional injection of energy. This energy is required to prevent gas cooling and therefore the runaway formation of stars \citep{2003ApJ...599...38B, 2006MNRAS.370..645B, 2006MNRAS.365...11C,2006Natur.442..888S}. As highly efficient sources of energy, accreting black holes at the centers of galaxies are now routinely invoked as the source of this energy and thus as a fundamental component in galaxy formation theory \citep{1998A&A...331L...1S} .

This process, called `feedback,' thus casts the black hole into the role of a thermostat for the gas in galaxies. By heating and expelling gas that would otherwise cool and condense into stars, black hole feedback is capable of fundamentally changing and controlling the evolutionary trajectory of their host galaxies and in turn the further growth of the black hole as it starves itself of fuel. This close relationship between galaxies and their central black holes can thus be described as `co-evolution,' potentially beginning with the birth of both in the early Universe.

\section{Evidence for Co-evolution}

The accretion of mass onto supermassive black holes and the conversion of baryons from gas into stars (observed as galaxy stellar mass growth -- star formation) follow similar general trends: the cosmic star formation history and the black hole accretion history track each other preserving roughly a ratio of 1000:1, which notably corresponds to the mass ratio between galaxy bulges and black holes observed in the local \Msigma\ relation \citep{2000ApJ...539L...9F, 2000ApJ...539L..13G, 2004ApJ...613..109H}, whose existence is also often invoked as evidence for co-evolution. Both star formation and black hole accretion history peak around $z\sim 1-2$ and then decline rapidly towards the present day.

The two growth histories also share further similarities: the most massive galaxies form in intense starbursts at high redshift while less massive galaxies have more extended star formation histories that peak later with decreasing mass \citep{1992MNRAS.254..601B, 2005ApJ...621..673T, 2010MNRAS.404.1775T, 2005ApJ...632..137N}. This `anti-hierarchical' nature is mirrored in black hole growth: the most massive black holes likely grow in intense quasar phases which peak in the early universe, while less massive black holes have more extended, less intense (low Eddington-ratio) growth histories that peak at lower redshift \citep{2003ApJ...598..886U,2005A&A...441..417H,2006AJ....131.2766R}

This does not imply that star formation rate and black hole accretion rate simply track each other within each galaxy; not every galaxy with a high star formation rate is also a quasar, and vice versa. Rather, within individual galaxies there seems to be an interplay, a co-evolution which regulates the whole galaxy--black hole system to conform to the general trend. How this regulation works is at the heart of the question of how co-evolution works.

The other major piece of evidence often cited in favor of co-evolution is the \Msigma\ relation; since the gravitational sphere of influence of the black hole is tiny compared to the host galaxy, fine tuning of the growth of both is required to produce a tight relation, as is observed. However, recent work by Peng(2007; \cite{2007ApJ...671.1098P}, refined by Jahnke \& Maccio \cite{2011ApJ...734...92J}) argues that a correlation between galaxy and black hole mass need not be the result of a causal relation at all. Rather, the nature of hierarchical assembly in a $\Lambda$CDM universe naturally results in a correlation after a sufficiently large number of mergers via the Central Limit Theorem. This does not mean that the \Msigma\ relation really has a non-causal origin, but rather that its interpretation is not straightforward.

\section{Multiple Modes of Co-evolution}

Recent observational advances in understanding the co-evolution of galaxies and black holes point to the existence of multiple modes. Thus, taking a `macro' view of both the galaxy and the active galactic nucleus (AGN) host galaxy population is critical  in any attempt to disentangle the large number of physical parameters such as mass, environment, morphology, and star formation rate. \textit{Large samples and high-quality multi-wavelength data are thus essential if we are to map out the evolutionary pathways that lead to co-evolution and feedback.}

In order to understand co-evolution, we need to take such samples and assess two fundamental questions. 

\begin{enumerate}
\item \textbf{At what stage in their lives do galaxies feed their black holes?}\\
What are the physical properties of galaxies where accretion is favored? Are these galaxies transitioning from one evolutionary stage to another, or are they representatives of a general, stable phase? How many different, separate AGN host galaxy populations are there, \textit{i.e.} how many pathways are there to black hole accretion?
\item \textbf{What effect does black hole growth have on the evolutionary trajectory of galaxies?}\\
This question must be answered separately for each AGN host galaxy population: does the energy liberated by the accretion phase actually impact the host galaxy, or does it dissipate without consequences? How does the energy couple to the gas in the host galaxy, and how does this depend on accretion mode (radiative vs. kinetic feedback)?
\end{enumerate}

It is important to address both questions as it is possible to imagine particular phases of galaxy evolution leading to accretion, but not feedback. In this case, the assumption that a high AGN fraction in this population is indicative of a high impact of AGN on the host galaxies is misleading.

\subsection{The Local Universe}

The advent of large samples from the Sloan Digital Sky Survey \citep{2000AJ....120.1579Y} has enabled large studies of AGN host galaxies. These studies showed that AGN host galaxies appear to be an `intermediate' population: they are neither blue and actively star-forming, nor are they red and passively evolving; rather, they reside in the `green valley' in between the two general galaxy populations. Their morphology also appeared to be in-between: large bulges with some disk component. These observations have led to the interpretation of AGN host galaxies as galaxies undergoing transformation from blue star-forming galaxies to red and dead ellipticals \citep{2003MNRAS.346.1055K}. 

Recently, a more complex picture has emerged. The local AGN host galaxy population is in fact a composite of two distinct groups. The fraction of the local AGN host galaxy in major mergers is negligible \citep{2010MNRAS.401.1552D}, though the incidence of mergers in the \textit{Swift} BAT sample is significantly higher \citep{2011ApJ...739...57K}; this is most likely due to the very shallow flux limit of the BAT sample which biases it to high luminosity objects. The remainder is divided between early-type galaxies ($\sim$10\%) and late-type galaxies ($\sim90\%$). Both of these two populations are very specific subsets of their parent populations (early-type and late-type galaxies, respectively) indicating that there are two fundamentally different modes of black hole fueling and co-evolution in the local universe (\cite{2010ApJ...711..284S} and Figure \ref{fig:gz_sdss}):

\begin{figure*}
\begin{center}

\includegraphics[width=0.99\textwidth]{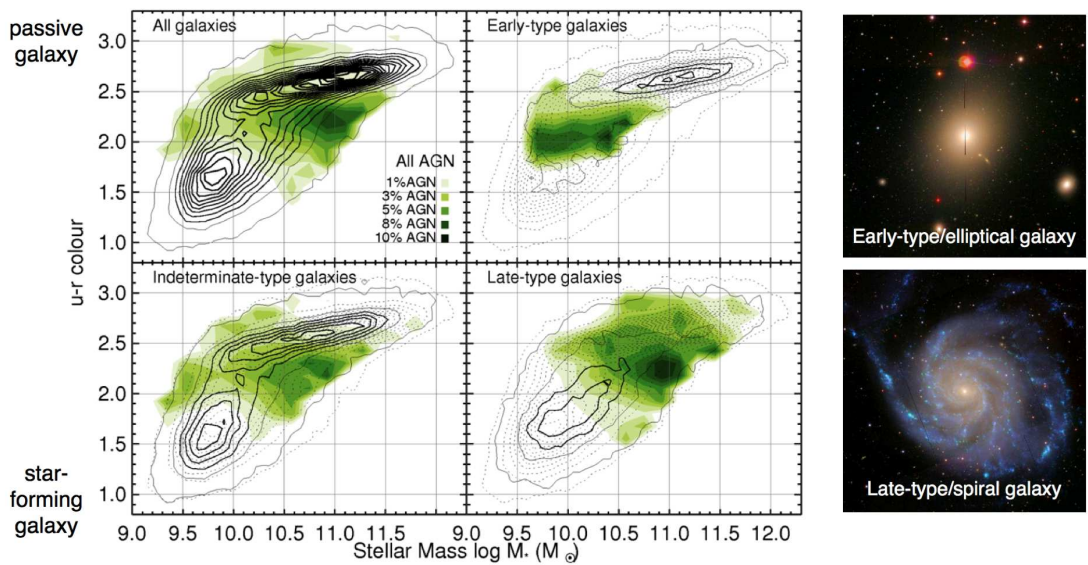}

\caption{The distribution of the fraction of galaxies that host AGN on the color-mass diagram together with example images of an early-type (or elliptical) galaxy (top right) and a late-type (or spiral) galaxy (bottom right). The contours represent the normal galaxy population while the green shaded contours trace the fraction of galaxies with growing central black holes. The active fraction in a specific sub-population is a proxy for the duty cycle of AGN in that population; it reveals which galaxy populations have a high black hole growth duty cycle and illustrates the importance of morphology for understanding the role of black hole growth in galaxy evolution \citep{2010ApJ...711..284S}.  The morphological classifications were made by the citizen scientists taking part in the Galaxy Zoo project (\texttt{galaxyzoo.org}, \citealt{2008MNRAS.389.1179L, 2011MNRAS.410..166L})}

\label{fig:gz_sdss}

\end{center}
\end{figure*}

\subsection{Early-type AGN Host Galaxies -- Merger-driven Migration from Blue to Red
}

In the early-type galaxy population, black hole growth occurs in the least massive ($10^{9.5-10.5}$\Msun) early-type galaxies with the bluest host galaxy colors amongst the early-type population. They reside in the `green valley,' and their stellar populations show that they feature post-starburst objects migrating from the blue cloud to the red sequence at roughly fixed stellar mass, thus populating the low-mass end of the red sequence \citep{2007MNRAS.382.1415S}. They \textit{may} represend a `downsized' version of the formation process undergone at high redshift by their more massive counterparts \citep{2005ApJ...621..673T, 2010MNRAS.404.1775T}. Deep imaging of early-type galaxies along the evolutionary sequence migrating from blue to red indicate that this migration is initiated in at least half, and perhaps all cases, by a merger or interaction \citep{2010ApJ...714L.108S}.

However, the concentration of AGN host galaxies in the green valley also challenges the traditional picture of the quenching of star formation by AGN feedback. Even assuming instantaneous quenching, stellar evolution dictates that the migration from the blue cloud to the green valley takes at least $\sim 100$ Myr (roughly the lifetime of OB stars). Detailed stellar population age-dating by Schawinski et al. (2007) \citep{2007MNRAS.382.1415S} shows that the typical post-starburst timing of high-Eddington Seyfert activity ranges from $\sim270$ Myr to $\sim1$ Gyr, \textit{i.e.} long after the quenching event. Thus, the black hole growth in the green valley and the associated energy that is liberated cannot be responsible for the shutdown of star formation. An unbiased search for AGN host galaxies using hard X-rays yields no `missing' powerful AGN in blue host galaxies \citep{2009ApJ...692L..19S}.

This observation does not entirely negate AGN as the agent responsible for the shutdown of star formation, it merely shows that the radiatively efficient Seyfert phase in the green valley is not the accretion phase responsible for the quenching of star formation. It may simply be the `mopping up' phase. At earlier times along the transition from the blue cloud to the red sequence, early-type galaxies do lose their molecular gas -- the fuel for star formation -- very rapidly \citep{2009ApJ...690.1672S}. This destruction of the molecular gas reservoir occurs before the high-Eddington Seyfert phase but coincides with weak AGN photoionization being present in the optical spectrum combined with still on-going star formation. Could this be the phase where a radiatively inefficient AGN is destroying the molecular gas reservoir?

Simple modeling of the depletion of molecular gas reservoirs following the Schmidt law for star formation \citep{1959ApJ...129..243S} by Kaviraj et al. (2011) \citep{2011MNRAS.415.3798K} shows that in the absence of an extra forcing mechanism, galaxies cannot rapidly quench their star formation since star formation is a self-regulated process. Every dynamical time, $t_{\rm dyn}$, they will convert a fraction of their available gas reservoir $M_{\rm gas}$ into stars with some efficiency $\epsilon$ (canonically $\sim0.02$) resulting in a depletion timescale for gas-rich galaxies of many Gyrs---precisely what is observed for star-forming spirals. In order to rapidly quench star formation and enable the migration to the red sequence within $\sim 1$ Gyr, some process beyond star formation alone must destroy or make unavailable the present molecular gas reservoir; the best candidate for this process is (kinetic) AGN feedback. 

\subsection{Late-type AGN Host Galaxies -- Secular Evolution and Stochastic Feeding
}

Most (up to 90\%) of local AGN host galaxies are massive spirals. They show no evidence for post-starburst stellar populations \citep{2011arXiv1111.1785W} or morphological disturbances indicating a recent catastrophic interaction or change in star formation rate. In fact, the typical late-type AGN host galaxy has the physical parameters of the Milky Way \citep{2010ApJ...711..284S,2011ApJ...736...84M}: a massive spiral with a low specific star formation rate---hence the green valley host galaxy colors, in contrast to the early-types whose color is due to post-starburst stellar populations. Since there is no evidence for significant external forcing of the system, the most likely explanation for the accretion seen in these late-type host galaxies is stochastic feeding of the black hole via secular processes \citep{2004ARA&A..42..603K}.

Since the typical local AGN host galaxy is similar to the Milky Way, the Galactic Center makes an excellent case study for what precisely leads to black hole feeding and feedback. While quiescent at the moment, observations show that the black hole in the Galactic Center was a low luminosity AGN as recently as 300 years ago as seen in hard X-ray light echos traveling across the molecular clouds surrounding the black hole \citep{2004A&A...425L..49R, 2008PASJ...60S.191N, 2011ApJ...739L..52N}. The recently-discovered Fermi Bubble may also be a remnant of recent accretion \citep{2010ApJ...724.1044S}.

\begin{figure*}
\begin{center}

\includegraphics[angle=90, width=0.99\textwidth]{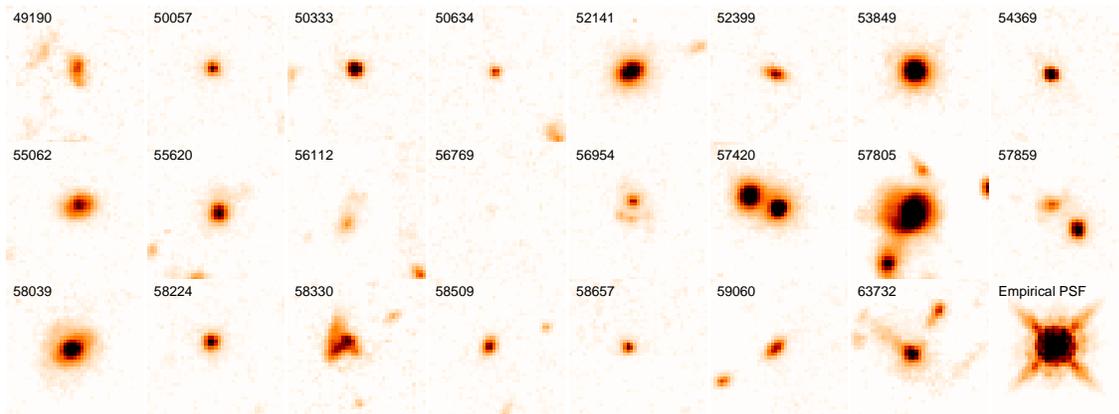}

\caption{\textit{Hubble Space Telescope} WFC3/IR images of typical z~2 moderate luminosity AGN host galaxies. These images are taken with the infrared channel of the Wide Field Camera 3 and show for the first time the rest-frame optical morphologies of $z\sim2$ AGN. Analysis of the surface-brightness profiles of these galaxies show disk-like profiles in 80\% of cases with the remainder composed of bulges and mergers. This means that a significant fraction of cosmic black hole growth in typical galaxies occurred in disk galaxies and are therefore likely driven by secular processes rather than major mergers as suggested by simulations by theorists \citep[][]{2011ApJ...727L..31S}.}

\label{fig:wfc3_agn}

\end{center}
\end{figure*}

\subsection{The High Redshift Universe}

The bulk of both black hole growth and star formation occurs at high redshift with the peak epoch for both occurring around $z\sim2$. This peak epoch has been difficult to study due to the lack of deep, high quality observations.  Deep X-ray surveys over the last decade have captured black hole growth out to very high redshift and down to relatively low luminosities and have revealed a large population of moderate-luminosity AGN at $1<z<3$ where normal mass black holes grow \citep[see reviews by][]{2005ARA&A..43..827B,2011arXiv1112.0320T}. This growth is slow as the Eddington ratios are moderate \citep{2011ApJ...734..121S}.  

Rest-frame optical spectroscopy from the ground is extremely challenging even with 8-10m telescopes \citep[e.g.][]{2009ApJ...706..535T} while space-based infrared spectroscopy is only now becoming possible with the new \textit{Hubble Space Telescope} Wide Field Camera 3 (WFC3) which has a slitless (grism) mode \citep[e.g.][]{2011AJ....141...14S, 2011ApJ...743L..37S}.

Recent imaging observations with WFC3/IR have revealed that the majority of the moderate luminosity AGN at $1<z<3$ feature disk light profiles rather than being massive bulges or major mergers (\citealt{2011ApJ...727L..31S} confirmed by \citealt{2011arXiv1109.2588K}). This implies that the high redshift AGN host galaxy population is in fact very similar to that in the nearby universe: mostly disk galaxies, a few spheroids and virtually no major mergers. Thus, black hole growth at $z\sim2$ is likely driven by stochastic fueling and secular processes.  Observations from \textit{Herschel} by \cite{2011MNRAS.tmp.1756M} support this picture as the far-infrared derived specific star formation rates of X-ray selected AGN host galaxies up to $z\sim3$ are indistinguishable from the underlying galaxy population.

Combining this observation with what we know from the local universe, we arrive at a picture where a significant fraction of cosmic black hole growth can be attributed to secular processes (\citealt{2011ApJ...727L..31S} estimate $\sim 40\%$) and that major mergers as a driver for black hole growth may be restricted to only a small fraction of the black hole growth in normal mass galaxies. Only at the highest luminosities do highly disturbed morphologies indicative of major mergers begin to appear \citep[e.g.][]{2008ApJ...674...80U,2006AJ....132.1496Z}.

A large caveat on any conclusions drawn from X-ray selected sample is that even the deepest \textit{Chandra} deep field surveys miss the most obscured black hole growth phases even when the underlying bolometric luminosity of the AGN is high. The X-ray emission from these AGN is so heavily absorbed that they are only betrayed by `excess' mid-infrared emission from hot dust in the nucleus, which even large amounts of extinction cannot remove \citep[e.g.][]{2004ApJ...616..123T, 2007ApJ...670..173D, 2008ApJ...672...94F,2009ApJ...706..535T,  2009ApJ...693..447F, 2010ApJ...722L.238T}. Expectations from simulations \citep[e.g.][]{2005Natur.433..604D, 2005ApJ...630..705H, 2005ApJ...625L..71H} indicate that heavily obscured, high luminosity AGN (quasars) at high redshift should reside in galaxies undergoing major mergers, and that it is during this obscured phase preceding the classical unobscured quasar phase that the quasar begins to blow the gas, thus quenching the star formation in the host galaxy. Near future observation with \textit{Hubble} may shed light on these poorly studied obscured AGN and whether they conform to the picture expected from simulations.

\begin{figure*}
\begin{center}

\includegraphics[width=0.49\textwidth]{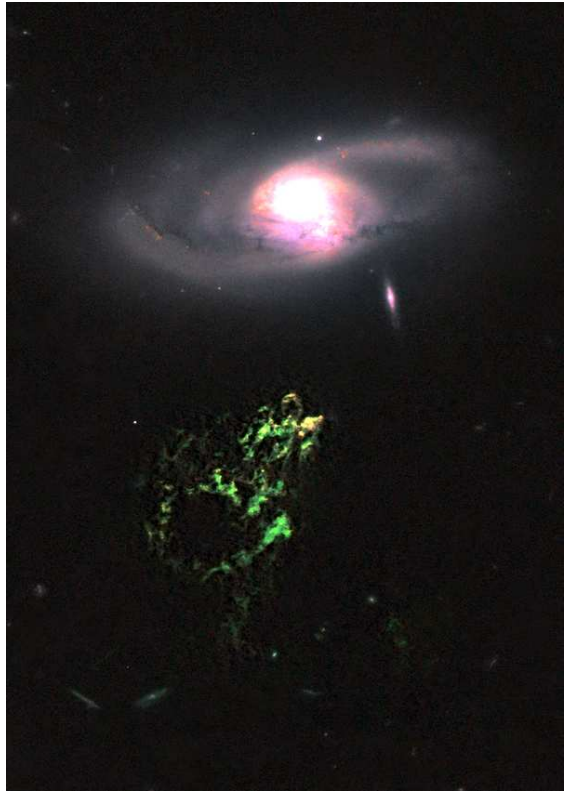}

\caption{\textit{Hubble Space Telescope} image of the galaxy IC 2497 (top) and Hanny's Voorwerp (green, [OIII] 5007). The Voorwerp is a light echo from a past quasar episode of the dormant black hole in IC 2497. The light travel time between the nucleus and the Voorwerp implies a time delay of less than 70,000 years, preserving a record of a powerful quasar phase in IC 2497 in the recent past (credit: NASA, ESA, W. Keel (University of Alabama), and the Galaxy Zoo Team) }

\label{fig:voorwerp_hst}

\end{center}
\end{figure*}

\section{Accretion States and Transitions}

The previous discussion of AGN host galaxy properties approaches the second question -- the question of whether the energy liberated by black hole growth actually affects the host galaxy -- only indirectly by observing that most AGN host galaxies appear to be stable disk galaxies with, at least locally, no evidence for recent enhancement or quenching of star formation. 

This raises the question of where we actually see AGN terminating, or at least modifying, star formation properties in their host galaxies. More or less the only unambiguous cases are the giant ionized outflows observed in powerful radio galaxies by \cite{2006ApJ...650..693N,2008A&A...491..407N}. Even in the case of dramatic examples such as the powerful molecular gas outflows seen in Mrk 231 by Fischer et 
al.(2010) \citep{2010A&A...518L..41F}, a starburst-driven wind cannot be entirely ruled out.

The commonality of black hole accretion phenomenology and physics exhibited by X-ray binaries (XRBs) and AGN has been discussed extensively in the literature \citep[e.g.][]{2003MNRAS.345L..19M, 2006Natur.444..730M, 2006MNRAS.372.1366K} and is particularly interesting since some XRBs are known to put out large amounts of kinetic energy, directly impacting their environment \citep[e.g.][]{2005Natur.436..819G, 2010Natur.466..209P}. If, as has been hypothesized, AGN can undergo rapid accretion state transitions similar to XRBs, could the radiatively efficient (quasar) phase be less important for feedback work? And could radiatively inefficient, kinetic outflows be how a large fraction of feedback work by black holes actually occurs? If so, then how do we find these radiatively inefficient---i.e. dim---AGN?

One case where this \textit{might} be occurring is the nearby ($z=0.05$) galaxy IC 2497, which features a light echo of a recent (>200,000 yr) powerful quasar outburst preserved on a giant external atomic hydrogen cloud \citep[][see Figure \ref{fig:voorwerp_hst}]{2009MNRAS.399..129L, 2009A&A...500L..33J, 2010A&A...517L...8R}. The nucleus of IC 2497, which hosted a $L_{\rm bol} \sim 10^{46-47}$\ergs\ less than 200,000 years ago, is now at least four orders of magnitude dimmer \citep{2010ApJ...724L..30S}. The rapidity of this `switch off' makes IC 2497 the best and most accessible place to test the hypothesis that quasars can undergo state transitions into a radiatively inefficient, kinetic mode. 

There are now other examples of similar 10-100 kyr timescale variability in local AGN \citep{2011MNRAS.tmp.2099K, 2011A&A...530A..60M,2011A&A...525A...6M}, giving further support to the XBR-AGN analogy and raising the prospect that radiatively inefficient accretion modes are an important missing piece in the feedback puzzle.

\section{Open Questions \& Prospects}

The role of black holes and their capacity to liberate large amounts of energy during their growth phases remains one of the thorniest issues in extragalactic astrophysics. While we are approaching a fairly complete census of black hole growth and host galaxy properties in the local universe, the high redshift universe remains only partially explored as low-luminosity and heavily-obscured AGN remain elusive in even the deepest surveys. Recent observational advances point to an increasing importance of secular processes in feeding black hole growth and to multiple, distinct pathways to black hole feeding. Thus, the answer to question 1 is that there are a number of very different points along the evolutionary pathways of galaxies during which black hole growth occurs.

The evidence for feedback remains conflicting and may well be limited to very specific phases of galaxy evolution, such as spheroid formation. The possibility that feedback is kinetic in form driven by radiatively inefficient black hole growth remains attractive. Despite this it remains poorly explored in the galaxy regime. Thus, question 2 remains more complex, though answers may be forthcoming in the near future. In particular, direct observations of black hole feedback on molecular gas reservoirs using the Atacama Large Millimeter Array (ALMA) will likely be revealing.

Present-day facilities such as \textit{Chandra} and \textit{Hubble}, combined with the next generation of near-infrared spectrographs on 8-10m telescopes, will continue to inform our picture of galaxy and black hole growth at high redshift as first glimpses of the earliest phases of both start to come into view \citep{2011Natur.474..356T}, though fully understanding the black hole -- galaxy connection all the way to the first objects in the universe may require the \textit{James Webb Space Telescope}.

\section*{Acknoweldgements}
I am grateful for the Bash `11 Symposium organizers at the University of Austin -- Texas for inviting me and for organizing such a wonderful meeting and I thank Ezequiel Treister and Meg Urry for comments. I acknowledge support by NASA through an Einstein Postdoctoral Fellowship (PF9-00069), issued by the Chandra X-ray Observatory Center, which is operated by the Smithsonian Astrophysical Observatory for and on behalf of NASA under contract NAS8-03060.

\end{document}